\documentclass[prl,twocolumn,aps,letterpaper, preprintnumbers,superscriptaddress]{revtex4}

\usepackage[english]{babel}

\usepackage{xr}
\usepackage{amsmath}   
\usepackage{amssymb}   

\usepackage{graphicx}
\usepackage[colorlinks=true, allcolors=blue]{hyperref}
\usepackage{float}
\usepackage{microtype}
\usepackage{ragged2e}
\usepackage{etoolbox}

\makeatletter
\renewcommand{\fnum@figure}{FIG.\ \thefigure}

\long\def\@makecaption#1#2{%
  \vskip\abovecaptionskip
  \begingroup
    \small
    \rmfamily
    \justifying
    \noindent #1.\ #2\par
  \endgroup
  \vskip\belowcaptionskip
}
\makeatother
\begin{document}

\title{Hopfion Superlattices in 3D Twisted Moir\'e Light Fields}

\author{Yifan Zhang}
\affiliation{School of Electrical and Electronic Engineering, Nanyang Technological University, Singapore 639798, Singapore}
\affiliation{Centre for Disruptive Photonic Technologies, School of Physical and Mathematical Sciences, Nanyang Technological University, Singapore 637371, Singapore}

\author{Vasu Dev}
\affiliation{Centre for Disruptive Photonic Technologies, School of Physical and Mathematical Sciences, Nanyang Technological University, Singapore 637371, Singapore}
\affiliation{Department of Physics, School of Advanced Sciences, Vellore Institute of Technology, Vellore 632014, Tamil Nadu, India.}

\author{Yijie Shen}
\email{yijie.shen@ntu.edu.sg}
\affiliation{Centre for Disruptive Photonic Technologies, School of Physical and Mathematical Sciences, Nanyang Technological University, Singapore 637371, Singapore}
\affiliation{School of Electrical and Electronic Engineering, Nanyang Technological University, Singapore 639798, Singapore}

\date{\today}

\begin{abstract}
Moir\'e superlattices have recently been extended from electronic materials to photonic systems, enabling complex topological light fields such as skyrmion bags. However, these realizations remain confined to two dimensions. Here, we demonstrate 3D moir\'e superlattices formed by twisting multi-volume spatiotemporal optical lattices, giving rise to previously undiscovered 3D quasiparticle topologies, including hopfion bags, hopfion quasicrystals, and hopfion-hopfionium nested superlattices. By tailoring the interference polarization textures of the moir\'e spacetime lattices, we achieve controlled topological transitions among these states, and tailored topological numbers of hopfion bags. These spatiotemporal supertopologies offer a promising platform for high-capacity information encoding in higher dimensions.
\end{abstract}

\maketitle

Moir\'e patterns are large-scale periodic superlattices formed by interfering two or more periodic structures with a relative twist or lattice mismatch. Originating in silk textiles, they were later exploited in metrological gratings for applications such as precision displacement sensing and strain analysis~\cite{Beranek1982,Post2008}. The field underwent a decisive turn when it was discovered that twisting two atomically thin material layers can fundamentally alter their physical properties, giving rise to the field of twistronics~\cite{PhysRevB.95.075420}. Research on twistronics has since produced numerous significant discoveries, including flat bands~\cite{Tian2023}, strong correlation~\cite{Campbell2024}, superconductivity~\cite{Cao2018b}, correlated insulator behavior~\cite{Cao2018}. This paradigm has also been extended to optics, where moir\'e superlattices of multilayer photonic crystals and materials provide new degrees of freedom for light manipulation~\cite{XIAO20201142,Hu2020,10.1021/acs.nanolett.0c01673,Oudich2024,Du2023Science,Wang2026,WU2025113833,PhysRevApplied,Luan2023,tang2026bilayer,tang2026moire}.
\begin{figure}[b]
\centering
\includegraphics[width=0.5\textwidth]{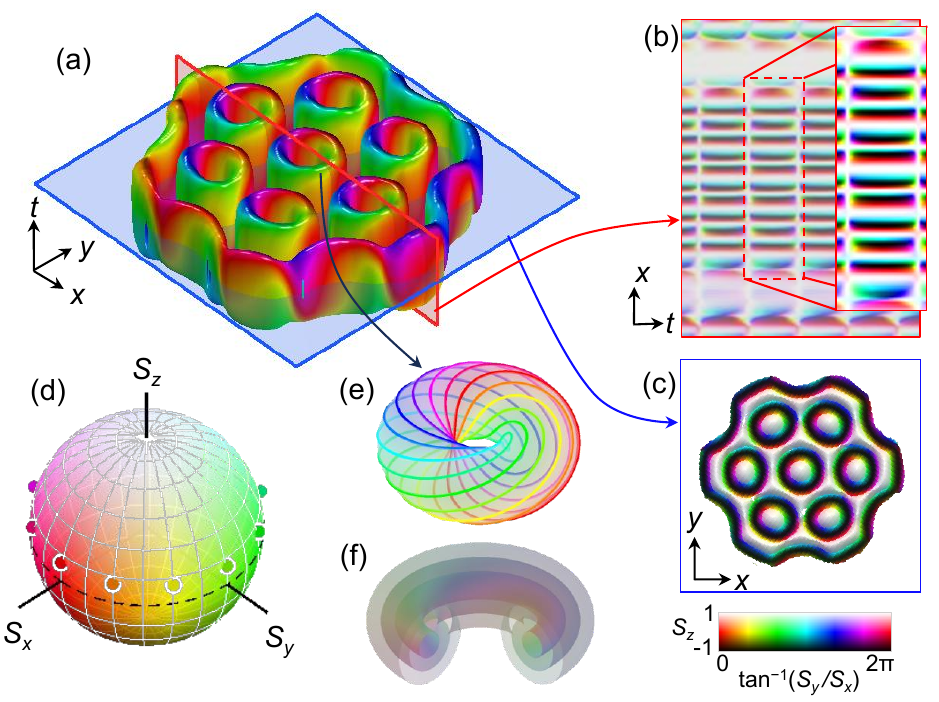}
    \caption{{(a)} Concept of a 3D hopfion bag comprising multiple elementary hopfions enclosed within a larger bag hopfion. {(b,c)} Polarization textures in the $xt$ and $xy$ planes, with skyrmion and skyrmionium bag topologies, respectively. {(d)} Selected polarization states on the Poincaré sphere and {(e)} their corresponding preimages forming linked closed loops in three-dimensional space-time. {(f)} Cutaway view of nested toroidal surfaces corresponding to circles of constant latitude on the Poincaré sphere, illustrating the Hopf-fibration structure.}
    \label{fig:1}
\end{figure}

Topological textures of light such as 2D skyrmions~\cite{Shen2024,shen2025free,cheng2025navigating}, 3D hopfions~\cite{shen2023,sugic2021particle,wang2023photonic,wu2025photonic}, and their assemblies of skyrmion or meron lattices~\cite{skyrmion2018,zhang2024periodic,wu2025optical,marco2024propagation,PhysRevLett.127.237403,wang2026optical} and recently extended hopfion crystals~\cite{hh5s-cprt}, have emerged as a powerful platform for multidimensional light manipulation with inherent stability immune to perturbations and capability of encoding data~\cite{peters2026extracting,wang2024topological,guo2026topological,chen2026programmable,zhang2026skyrmions}. Recently, the twistronics paradigm has been extended to topological light waves. The twisted bilayer plasmonic skyrmion lattices can generate moir\'e superlattices hosting skyrmion bags~\cite{Schwab2025,trilayer2025} and skyrmion clusters~\cite{zhang2025observation}. The upgraded twisted trilayer skyrmion bags show stronger topological robustness~\cite{trilayer2025}. Twisted moir\'e topologies were also studied in photon polariton~\cite{mangold2026phonon,tunable2026}, optical spin~\cite{Shi2026}, as well as freespace structured light enabling propagating moir\'e supertopologies~\cite{dev}. While 2D topological textures have been extensively explored, extending topological structures to 3D remains a challenge for accessing richer and intrinsically higher-dimensional topologies with moir\'e synergy.

In this Letter, we demonstrate a method to generate 3D spatiotemporal moir\'e superlattices with 3D topologies. Using a bichromatic light field, we evolve 2D skyrmionium lattices along the temporal dimension into 3D hopfion crystals, then superimpose two hopfion crystals with a twist angle to generate hopfion-based moiré superlattices. In which, the supercell manifests diverse novel topologies such as, hopfion bags, isolated hopfions, nested superlattices, and quasicrystals, thereby filling the gap of 3D topological moir\'e superlattices. 
As the basic supercell structure of the periodic superlattice, the hopfion bag represents a distinctive 3D spatiotemporal topology, in which multiple elementary hopfions are enclosed within a larger outer hopfion with opposite winding, as shown in Fig.~\ref{fig:1}{\color{blue}(a)}. In Figs.~\ref{fig:1}{\color{blue}(b,c)} we can see the polarization textures of the hopfion bag periodically repeat along the time dimension while revealing a large outer skyrmionium enclosing multiple inner skyrmioniums in $xy$ plane. Each sub-hopfion in the hopfion bag can be characterized by Hopf fibration structure [Fig.~\ref{fig:1}{\color{blue}(d-f)}]. These 3D moiré superlattices may thus offer enhanced topological stability and support higher-density information encoding through their nested, multilayered structures.

\begin{figure}[b]
    \centering
    \includegraphics[width=0.5\textwidth]{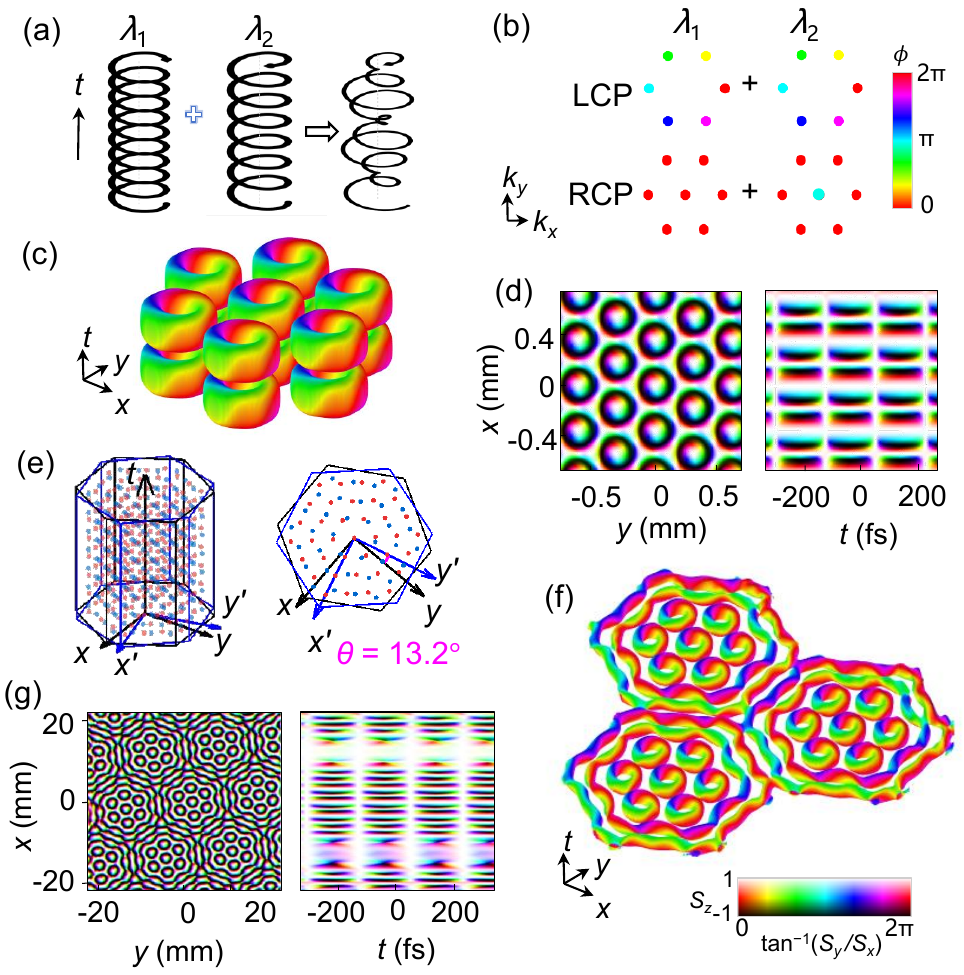}
    \caption{{(a)} Temporal beating arising from the superposition of two optical fields with distinct wavelengths $\lambda_1$ and $\lambda_2$. {(b)} Arrangement of bichromatic light sources in $k$ space for the left- and right-circularly polarized components. {(c)} Hopfion crystal generated from a single $C_6$-symmetric array of light sources. {(d)} Polarization textures of the hopfion crystal in the $xy$ plane at $t=0$ and the $xt$ plane at $y=0$, exhibiting skyrmionium and skyrmion topologies, respectively. {(e)} Schematic of two identical hopfion crystals superposed with a relative twist angle $\theta=13.2^\circ$, along with their transverse projection. {(f)} Three-dimensional spatiotemporal moiré superlattice of hopfion bags.  {(g)} Corresponding polarization textures of the hopfion-bag superlattice, revealing a periodic arrangement of skyrmionium bags in the $xy$ plane and skyrmion textures in the $xt$ plane.}
    \label{fig:2}
\end{figure}
The superposition of two coherent light fields with different polarization wavelengths $\lambda_1$ and $\lambda_2$ induces an optical frequency beat, with the corresponding pseudospin tracing a trajectory in spatiotemporal domain, as shown in Fig.~\ref{fig:2}{\color{blue}(a)}. The polarization evolution is governed by the relative phase $\Delta\omega t$, where $\Delta\omega=|\omega_1-\omega_2|$, and is periodic with period $T=2\pi/\Delta\omega$.
To extend the moiré superlattices from two spatial dimensions to 3D spatiotemporal topology, we exploit optical beat-frequency interference in topological lattice. The electric field of two superposed coherent fields with different wavelengths $\lambda_1$ and $\lambda_2$ can be expressed as:
\begin{equation}
\boldsymbol{E}(t)=\boldsymbol{E}_{1}\mathrm{e}^{-\mathrm{i}\omega_1 t}+\boldsymbol{E}_{2}\mathrm{e}^{-\mathrm{i}\omega_2 t},
\end{equation}
where $\omega_{1,2}=2\pi c/\lambda_{1,2}$. In our work, the two wavelengths are set to $\lambda_1=532\,\mathrm{nm}$ and $\lambda_2=1.01\lambda_1\approx537.32\,\mathrm{nm}$. To realize the 3D photonic lattice, we use a hexagonal array of light sources at near field [Fig.~\ref{fig:2}{\color{blue}(b)}], then the $N$-wave interference produces the underlying periodic lattice. The superposed sources employ orthogonal left- and right-handed circular polarizations (RCP \& LCP). The LCP sources carry phases spanning $0$ to $2\pi$ and contain two wavelength components, $\lambda_1$ and $\lambda_2$. The total LCP field can be expressed as the superposition of the fields from two wavelengths:
\begin{equation}
\begin{split}
\boldsymbol{E}_L={} & \sum_{m=1}^{n} G_m(\lambda_1) \exp\left( \mathrm{i} \frac{2\pi l m}{n} \right) \hat{\mathbf{e}}_L \\
& + \sum_{m=1}^{n} G_m(\lambda_2) \exp\left( \mathrm{i} \frac{2\pi l m}{n} \right)\hat{\mathbf{e}}_L.
\end{split}
\label{eq:EL}
\end{equation}
Here, $l=1$ denotes the discrete vortex charge imposed on the LCP component, introducing an azimuthal phase increment of $\pi/3$ between adjacent light sources. The RCP field is composed of six hexagonal  lightsources with another tunable source at the center, each contributing at both wavelengths $\lambda_1$ and $\lambda_2$. The total RCP field is given by
\begin{equation}
\begin{split}
\boldsymbol{E}_R ={} & \sum_{m=1}^{n} G_m(\lambda_1) \hat{\mathbf{e}}_R
      + \sum_{m=1}^{n} G_m(\lambda_2) \hat{\mathbf{e}}_R \\
      & + A_1 e^{\mathrm{i}\phi_1} G_c(\lambda_1) \hat{\mathbf{e}}_R
      + A_2 e^{\mathrm{i}\phi_2} G_c(\lambda_2) \hat{\mathbf{e}}_R,
\end{split}
\label{eq:ER}
\end{equation}
where $n$ denotes the number of light sources. The hexagonal lattice with \(C_6\) symmetry is formed by six interfering waves (\(n=6\)). 
 The amplitude profile of the $m^{th}$ source is described by a Gaussian function,
$G_m=\exp\left[-\frac{(x-x_m)^2+(y-y_m)^2}{w_0^2}\right],$
where $(x_m,y_m)$ specifies the transverse position of the source and $w_0$ is the beam-width. The central source profile amplitude is described by $G_c=\exp\left[-\frac{x^2+y^2}{w_0^2}\right]$. The maximum central-source  amplitudes are \(A_1\) and \(A_2\), with phases \(\phi_1=0\) and \(\phi_2=\pi\) at wavelengths \(\lambda_1\) and \(\lambda_2\), respectively. The total time-dependent optical field can be written as
\begin{equation}
\begin{split}
\boldsymbol{E}(x,y,t)
={}&
\boldsymbol{E}_{L}(x,y,\lambda_1)\mathrm{e}^{-\mathrm{i}\omega_1 t}
+
\boldsymbol{E}_{L}(x,y,\lambda_2)\mathrm{e}^{-\mathrm{i}\omega_2 t}
\\
&+
\boldsymbol{E}_{R}(x,y,\lambda_1)\mathrm{e}^{-\mathrm{i}\omega_1 t}
+
\boldsymbol{E}_{R}(x,y,\lambda_2)\mathrm{e}^{-\mathrm{i}\omega_2 t}.
\end{split}
\label{eq:Et}
\end{equation}
The resultant polarization texture is characterized by a generalized unit spin vector \(\boldsymbol{S}=\frac{\boldsymbol{E}^{\dagger}\boldsymbol{\sigma}\boldsymbol{E}}{\boldsymbol{E}^{\dagger}\boldsymbol{E}}\), where \(\boldsymbol{\sigma}=(\sigma_x,\sigma_y,\sigma_z)\) denotes the vector of Pauli matrices. This unit vector characterizes the local polarization state and maps each point in the \((x,y,t)\) domain onto the Poincaré sphere, providing the basis for characterizing the hopfion topology.

The Fourier transform of the source array in $k-$space, combined with bichromatic interference, generates a spatiotemporal hopfion crystal exhibiting periodicity in both space and time, the lattice structure is presented in Fig.~\ref{fig:2}{\color{blue}(c)}. In the $xy$ corss-section of $t=0$, the polarization texture forms a hexagonal $C_6$ lattice of skyrmioniums. The interference between the two frequency components introduces temporal periodicity, giving rise to a periodic skyrmion lattice in the $xt$ plane at $y=0$, see Fig.~\ref{fig:2}{\color{blue}(d)}. A hopfion crystal can also be generated using four light sources arranged with $C_4$ symmetry ($n=4$), as shown in Supplemental Material, {\color{blue}Sec. S1}.

To realize a 3D moiré superlattice, we superpose two identical hopfion crystals with a relative twist angle $\theta$ about the lattice center, chosen to satisfy the commensurability condition. The resultant configuration is schematically shown in Fig.~\ref{fig:2}{\color{blue}(e)}, where the blue and red dots represent the original and rotated hopfion crystals, respectively, in $(x,y,t)$ space. At $\theta=13.2^\circ$, interference between the constituent fields generates a spatiotemporal moiré superlattice of hopfion bags, see Fig.~\ref{fig:2}{\color{blue}(f)}. Each hopfion bag consists of several sub-hopfions enclosed within a larger bag hopfion of opposite Hopf index. The corresponding cross sections in Fig.~\ref{fig:2}{\color{blue}(g)} show lattice of skyrmionium bags in the $xy$ plane at $t=0$ and skyrmion textures in the $xt$ plane at $y=0$.

A hopfion is a  three-dimensional topological solition that fulfills the Hopf map \(\pi_3(S^2)\) with Hopf index
(\(Q_H\)), represented through a stereographic projection from \(S^3\) to \(\mathbb R^3\) (\(S^3\setminus\{0\}\cong\mathbb R^3\)). To quantitatively characterize the topology of the three-dimensional hopfion lattices and superlattices constructed in our work, we calculate the $Q_H$, a topological invariant that characterizes the global topology of the pseudospin texture. It is expressed in terms of the emergent magnetic field \(\mathbf{B}\) and its vector potential \(\mathbf{A}\) as~\cite{Whitehead1947,Linking},
\begin{equation}
Q_H =
\frac{1}{(4\pi)^2}
\iiint_{\Omega}
\mathbf{B}\cdot\mathbf{A}\,
dx\,dy\,dt ,
\end{equation}
here $\mathbf{B}$ is constructed from the normalized pseudospin texture $\mathbf{S}=(S_x,S_y,S_z)$, with its components defined as $B_i=\frac{1}{2}\epsilon_{ijk}\mathbf{S}\cdot(\partial_j\mathbf{S}\times\partial_k\mathbf{S})$, where $i,j,k\in\{x,y,t\}$ and $\epsilon_{ijk}$ is the Levi--Civita tensor. The vector potential $\mathbf{A}$ is defined by $\nabla\times\mathbf{A}=\mathbf{B}$, and $\Omega$ denotes the finite spatiotemporal integration domain. The detailed method for calculating the $Q_H$ value and Hopf density distribution of the superlattices is presented in {\color{blue}Sec. S2} of the Supplemental Material.

\begin{figure}[htbp]
    \centering
  \includegraphics[width=0.5\textwidth]{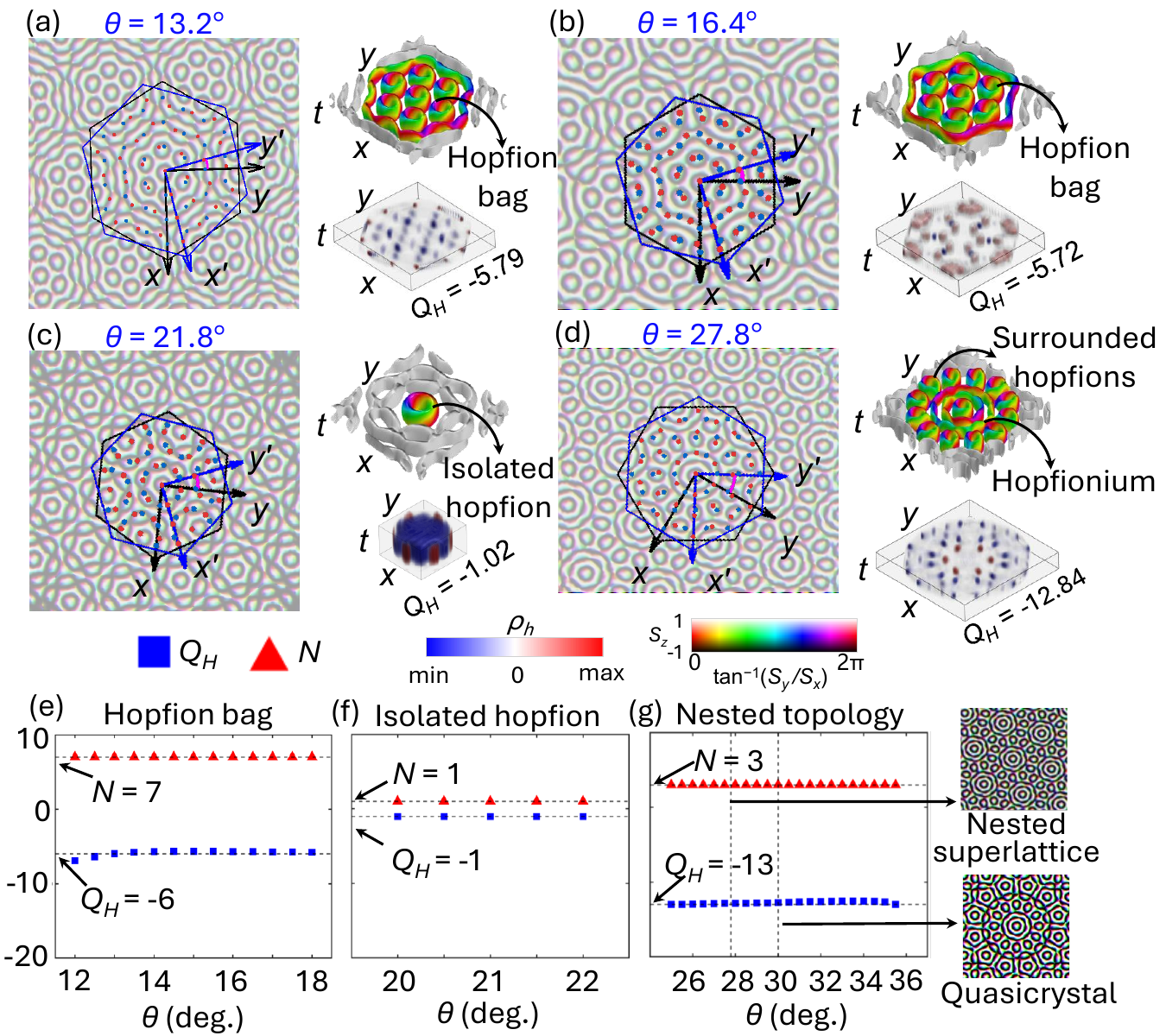}
    \caption{Twisting two lattice layers generates moir\'e interference superlattices,with their hopfion density distributions plotted. {(a)} At $\theta = 13.2^\circ$, a hopfion bag containing seven sub-hopfions emerges in spacetime. {(b)} At $\theta = 16.4^\circ$, the structure remains a hopfion bag. {(c)} At $\theta = 21.8^\circ$, isolated hopfion forms. {(d)} At $\theta = 27.8^\circ$, a nested superlattice appears. Evolution of $Q_H$ (blue) and sub-hopfion number $N$ (red) with twist angle $\theta$. {(e)} For $\theta \in [12^\circ, 18^\circ]$: hopfion bag, $N = 7$, $Q_H \approx -6$. {(f)} For $\theta \in [20^\circ, 22^\circ]$: isolated hopfion, $N = 1$, $Q_H \approx -1$. {(g)} For $\theta \in [25^\circ, 35.5^\circ]$: nested topology, $N = 3$, $Q_H \approx -13$. At $\theta = 27.8^\circ$: nested superlattice; at $\theta = 30^\circ$: quasicrystal.}
    \label{fig:3}
\end{figure}
Based on the $C_6$ spatiotemporal hopfion crystal, we twist the two lattice layers in $k$ space to generate moiré superlattices with different topological configurations. At twist angle  $\theta=13.2^\circ$, a hopfion bag forms with $Q_H=-5.79$, consisting of a large outer hopfion enclosing seven sub-hopfions of opposite polarity with the corresponding sub-hopfions number $N=7$, see Fig.~\ref{fig:3}{\color{blue}(a)}. Increasing the twist angle to $\theta=16.4^\circ$ modifies the outer hopfion shape while preserving the seven internal sub-hopfions [Fig.~\ref{fig:3}{\color{blue}(b)}], then $Q_H$ equals to -5.79. At $\theta=21.8^\circ$, the structure evolves into an isolated hopfion with $Q_H=-1.02$ and $N=1$ [Fig.~\ref{fig:3}{\color{blue}(c)}]. At $\theta=27.8^\circ$, a nested superlattice emerges, with each unit cell containing a radially nested hopfionium of three hopfions ($N=3$), surrounded by twelve peripheral hopfions [Fig.~\ref{fig:3}{\color{blue}(d)}], the total $Q_H$ equals to -12.84. The corresponding Hopf index of these superstructures are determined from the complete topological structure within one superlattice unit cell.

To investigate the evolution of the topological superstructures with twist angle, we continuously vary $\theta$ from $12^\circ$ to $35.5^\circ$ and track the number of sub-hopfions $N$ and the Hopf index $Q_H$ within a moiré unit cell. As shown in Fig.~\ref{fig:3}{\color{blue}(e–g)}, the system evolves from hopfion bags at $\theta=12^\circ$–$18^\circ$, through isolated hopfions at $\theta=20^\circ$–$22^\circ$, to nested hopfionium textures at $\theta=25^\circ$–$35.5^\circ$. The corresponding Hopf index remains close to $Q_H=-6$ and $Q_H=-1$ for hopfion bags and isolated hopfions, respectively. At higher twist angles, a periodic hopfionium superlattice emerges at $\theta=27.8^\circ$, evolving into a hopfionium quasicrystal at $\theta=30^\circ$. We further examine the propagation stability of these structures over $1$–$2$ wavelengths along $z$ (Supplemental Material, {\color{blue}Sec.~S3}).

\begin{figure}
    \centering
   \includegraphics[width=1\linewidth]{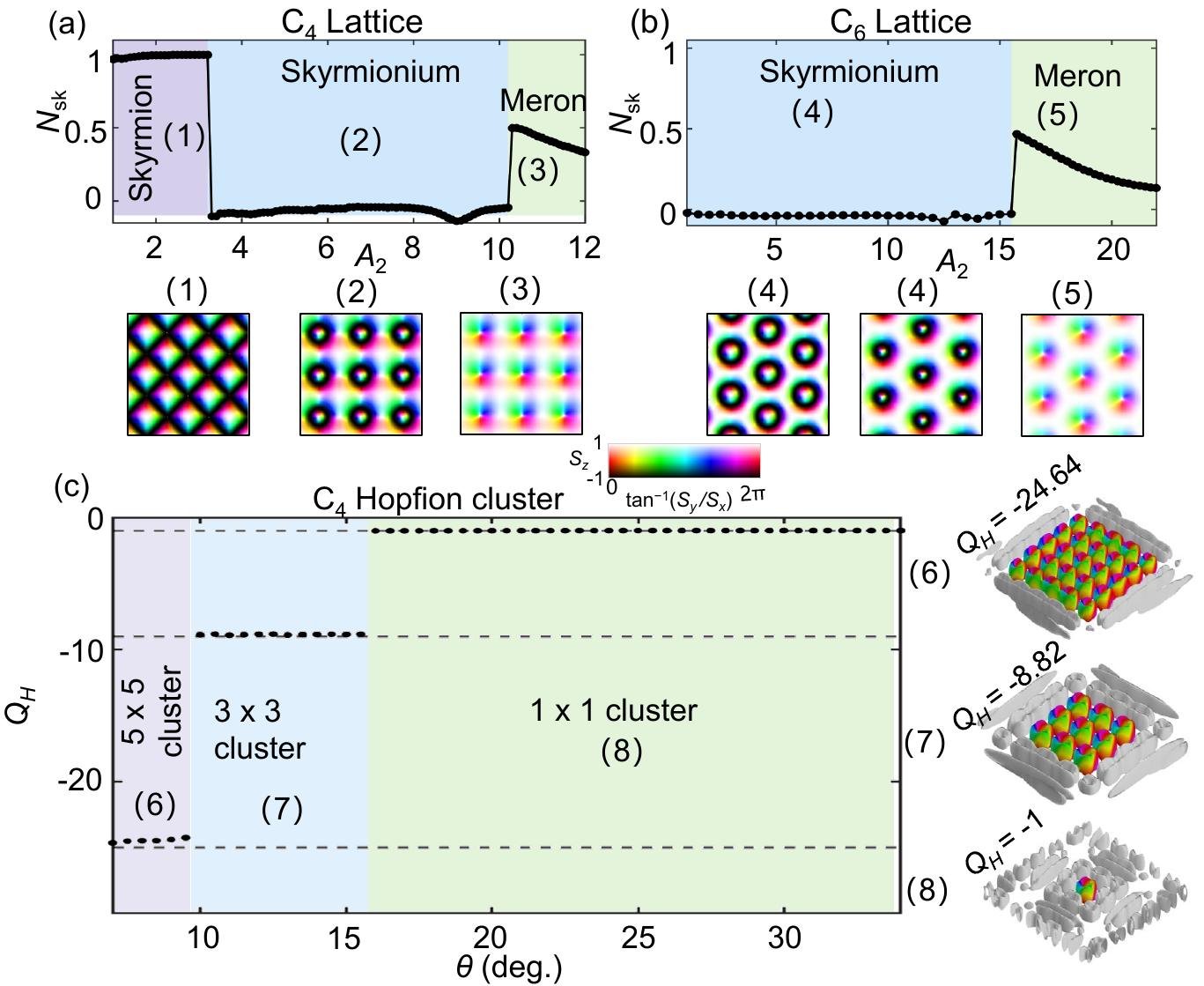}
    \caption{Topological transitions as a function of the amplitude of the central source for wavelength 537.32 nm in $C_4$ and $C_6$ lattice. {(a)} While $A_2$ changing from 1 to 12,the $C_4$ lattice topological pattern in 2D cross-section transits from (1) skyrmion to (2) skyrmionium  and then (3) meron. {(b)} In $C_6$ lattice , while the $A_2$ changing from 1 to 20, the topological pattern in 2D cross-section transits from (4) skyrmion to meron. {(c)} Meanwhile, rotating the two $C_4$ lattice in different angle $\theta$ causes the twisting superlattice transits from (6) $5\times5$ hopfion cluster to (7) $3\times3$ hopfion cluster and finally becomes (8) $1\times1$ hopfion cluster while $Q_H$ changing. }
    \label{fig:4}
\end{figure}
The effect of the central light-source amplitude on the topological states of the lattices is further investigated. We set $n = 6$ and $n = 4$, corresponding to the hexagonal and square symmetric arrangements of the source array, respectively. In our study, we fix the amplitude of the $\lambda_1$ source at unity ($A_1 = 1$), with relative phases $\phi_1=0$ and $\phi_1=\pi$. In the $C_4$ lattice, by tuning $A_2$ (for the $\lambda_2$ source) from $1$ to $12$, we observe the variation in the topological texture, as shown in Fig.~\ref{fig:4}{\color{blue}(a)}. For $n=4$, the polarization texture at $t=0$ forms a square lattice of skyrmions, each carrying a skyrmion number $N_{\mathrm{sk}}=1$. As $A_2$ increases, the skyrmion number decreases to $N_{\mathrm{sk}}=0$, marking a transition from skyrmions to skyrmioniums. A further increase in $A_2$ transforms the skyrmioniums into merons, with $N_{\mathrm{sk}}$ approaching $0.5$. Similarly, for the $C_6$ lattice, varying the amplitude $A_2$ from $1$ to $20$ results in a nonmonotonic variation in the skyrmion number $N_{\mathrm{sk}}$. Initially, $N_{\mathrm{sk}}$ increases from $0$ to approximately $0.5$, corresponding to a transformation of the polarization texture from a skyrmionium to a meron topology in the $xy$ plane at $t=0$ [Fig.~\ref{fig:4}{\color{blue}(b)}]. Meron textures emerge at $A_2=10.5$ and $15.1$ for the $C_4$ and $C_6$ lattices, respectively. Beyond these values, the increasing amplitude of the RCP component leads to a dominant RCP background, progressively suppressing the meron texture, as reflected by the decreasing $N_{\mathrm{sk}}$. Amplitude tuning thus provides a means of controlling structural transitions not only in these 3D lattices but also in the superlattices (see {\color{blue}Sec. S4} of the Supplemental Material). Besides, the relative twisting of two $C_4$-symmetric hopfion crystals results in hopfion clusters whose size depends on the twist angle $\theta$. As $\theta$ increases from $10^\circ$ to $30^\circ$, the clusters evolve from $5\times5$ to $3\times3$ and ultimately to $1\times1$ hopfion arrangements [Fig.~\ref{fig:4}{\color{blue}(c)}]. The corresponding Hopf index $Q_H$ changes from $-25$ to $-9$ and finally to $-1$. 

\begin{figure}
    \centering
\includegraphics[width=1\linewidth]{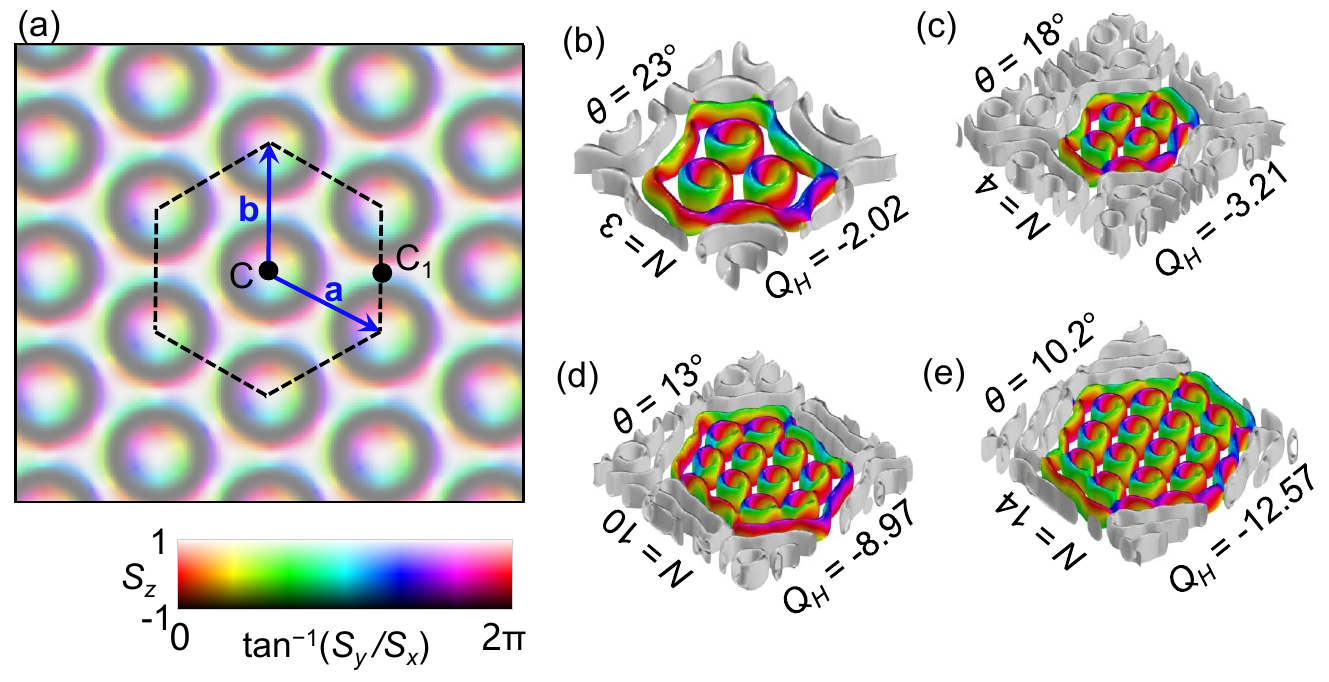}
    \caption{{(a)} Hopfion bags with tunable numbers of sub-hopfions, obtained by shifting the rotation center from the lattice center $C$ to the symmetry position $C_1$ at $\mathbf a+\frac{1}{2}\mathbf b$. {(b-e)} Three-dimensional hopfion bags containing 3, 4, 10, and 14 sub-hopfions,with corresponding $xy$-cross-sectional polarization patterns.}
    \label{fig:5}
\end{figure}
In addition, we control the number of sub-hopfions within the hopfion bag by changing the center location and angle of the symmetric rotation, specifically shifting the rotation centers of the two lattices from the center of the lattice unit $C$ to the position $C_1$ [Fig.~\ref{fig:5}{\color{blue}(a)}]. Then a hopfion bag containing three sub-hopfions is formed at a twist angle $\theta=$ 23$^\circ$ [Fig.~\ref{fig:5}{\color{blue}(b)}]. At a twist angle of 10.2$^\circ$ [Fig.~\ref{fig:5}{\color{blue}(c)}], a hopfion bag containing fourteen sub-hopfions can be obtained. Furthermore, by changing the rotation angle to other specific values, hopfion bags with 4, 10, or other numbers of sub-hopfions can be generated [Fig.~\ref{fig:5}{\color{blue}(d,e)}].

In summary, we have introduced a method for generating 3D spatiotemporal moiré superlattices through the superposition of relatively twisted bichromatic optical fields. Our approach realizes, for the first time, hopfion bags, isolated hopfions, and other complex moir\'e superlattices, including nested hopfionium in superlattices and quasicrystals — all genuine 3D topological structures, offering significantly higher information density and enhanced topological stability. Moreover, our structures are generated solely through the interference and twisting of light fields, without relying on any 2D material platforms. This material-free approach circumvents the issues of defects, scattering losses, and fabrication complexity inherent in material-based moir\'e systems, while also substantially improving the reconfigurability and programmability of the topological structures. We anticipate that the principles demonstrated here can be extended beyond optical fields to magnetic, acoustic, elastic, and water-wave systems, enabling the exploration of high-dimensional and spatiotemporal moiré superlattices across diverse physical platforms~\cite{CrX3,sound,Ochoa2021DegradationOP,Wang2025}. More broadly, our work establishes a new paradigm for high-dimensional topological moiré structures, with potential to enable robust, high-capacity information storage, transmission, and processing.

\textit{Acknowledgments} -- Authors thank Dr. Wenbo Lin for useful discussions and valuable suggestions. 

\textit{Funding} -- Singapore Ministry of Education (MOE) (RG157/23 \& RT11/23), Singapore Agency for Science, Technology and Research (A*STAR) (M24N7c0080 \& H25-MRO3489), Nanyang Technological University (NAP-SUG), Vellore Institute of Technology (VIT) Vellore (RGEMS, Sanction Order No. SG20260317).

\bibliographystyle{naturemag}
\bibliography{sample}

@article{tang2026moire,
  title={Moir{\'e} nanophotonics: from fundamental concepts to device engineering},
  author={Tang, Haoning and Ni, Xueqi and Zalogina, Anastasiia and Quard, Hugo and Coste, Nathan and Aharonovich, Igor},
  journal={Optica},
  volume={13},
  number={6},
  pages={1007--1019},
  year={2026},
  publisher={Optica Publishing Group}
}

@article{tang2026bilayer,
  title={Bilayer nonlocal flat optics},
  author={Tang, Haoning and Fan, Shanhui and Jie, Yao and Nguyen, Hai Son and Mazur, Eric and Cao, Yuan},
  journal={eLight},
  volume={6},
  number={1},
  pages={21},
  year={2026},
  publisher={Springer}
}

@article{Tian2023,
  author = {Tian, Haidong and Gao, Xueshi and Zhang, Yuxin and Che, Shi and Xu, Tianyi and Cheung, Patrick and Watanabe, Kenji and Taniguchi, Takashi and Randeria, Mohit and Zhang, Fan and Lau, Chun Ning and Bockrath, Marc W.},
  title = {Evidence for {Dirac} flat band superconductivity enabled by quantum geometry},
  journal = {Nature},
  volume = {614},
  number = {7948},
  pages = {440--444},
  year = {2023},
  doi = {10.1038/s41586-022-05576-2}
}

@article{PhysRevLett.127.237403,
  title = {Photonic Spin Lattices: Symmetry Constraints for Skyrmion and Meron Topologies},
  author = {Lei, Xinrui and Yang, Aiping and Shi, Peng and Xie, Zhenwei and Du, Luping and Zayats, Anatoly V. and Yuan, Xiaocong},
  journal = {Phys. Rev. Lett.},
  volume = {127},
  issue = {23},
  pages = {237403},
  numpages = {6},
  year = {2021},
  month = {Dec},
  publisher = {American Physical Society},
  doi = {10.1103/PhysRevLett.127.237403}
}

@article{mangold2026phonon,
  title={Phonon-polaritonic skyrmions: transition from bubble-to N{\'e}el-type},
  author={Mangold, Florian and Ba{\`u}, Enrico and Nan, Lin and Schwab, Julian and G{\"o}lz, Thorsten and Mancini, Andrea and Frank, Bettina and Tittl, Andreas and Giessen, Harald},
  journal={Light: Science \& Applications},
  volume={15},
  number={1},
  pages={239},
  year={2026},
  publisher={Nature Publishing Group UK London}
}

@article{wang2026optical,
  title={Optical quasisymmetry groups for meron lattices},
  author={Wang, Guangfeng and Feng, Juan and Zheng, Tong and Shen, Yijie and Chen, Xianfeng and Wang, Bo},
  journal={Physical Review Research},
  volume={8},
  number={3},
  pages={033226},
  year={2026},
  publisher={APS}
}

@article{wu2025optical,
  title={Optical skyrmion lattices accelerating in a free-space mode},
  author={Wu, Haijun and Zhou, Weijie and Zhu, Zhihan and Shen, Yijie},
  journal={APL Photonics},
  volume={10},
  number={5},
  year={2025},
  publisher={AIP Publishing}
}

@article{marco2024propagation,
  title={Propagation-invariant optical meron lattices},
  author={Marco, David and Herrera, Isael and Brasselet, Sophie and Alonso, Miguel A},
  journal={ACS Photonics},
  volume={11},
  number={6},
  pages={2397--2405},
  year={2024},
  publisher={ACS Publications}
}

@article{wu2025photonic,
  title={Photonic torons with 3D topology transitions and tunable spin monopoles},
  author={Wu, Haijun and Mata-Cervera, Nilo and Wang, Haiwen and Zhu, Zhihan and Qiu, Chengwei and Shen, Yijie},
  journal={Physical Review Letters},
  volume={135},
  number={6},
  pages={063802},
  year={2025},
  publisher={APS}
}

@article{wang2023photonic,
  title={Photonic spin hopfions and monopole loops},
  author={Wang, Haiwen and Fan, Shanhui},
  journal={Physical Review Letters},
  volume={131},
  number={26},
  pages={263801},
  year={2023},
  publisher={APS}
}

@article{shen2025free,
  title={Free-space topological optical textures: tutorial},
  author={Shen, Yijie and Wang, Haiwen and Fan, Shanhui},
  journal={Advances in Optics and Photonics},
  volume={17},
  number={2},
  pages={295--374},
  year={2025},
  publisher={Optica Publishing Group}
}

@article{cheng2025navigating,
  title={Navigating optical skyrmions—from historical origins to applications: tutorial},
  author={Cheng, Cheng and Rao, Lixi and Ye, Junyi and Zhao, Xingqi and Che, Zhiyuan and Liu, Wenzhe and Wang, Jiajun and Shi, Lei},
  journal={Advances in Optics and Photonics},
  volume={18},
  number={1},
  pages={1--105},
  year={2025},
  publisher={Optica Publishing Group}
}

@article{zhang2025observation,
  title={Observation of moir{\'e} plasmonic skyrmion clusters},
  author={Zhang, Lan and Wan, Lipeng and Deng, Weimin and Hou, Liang and Qiu, Jumin and Zou, Qiushun and Wang, Tongbiao and Zhao, Daomu and Yu, Tianbao},
  journal={Science Advances},
  volume={11},
  number={51},
  pages={eadx0478},
  year={2025},
  publisher={American Association for the Advancement of Science}
}

@article{zhang2026skyrmions,
  title={Skyrmions based on optical anisotropy for topological encoding},
  author={Zhang, Yunqi and Wang, An Aloysius and Zhang, Runchen and Zhao, Zimo and Ma, Yifei and Liu, Ruofu and Pong, Zhi-Kai and Cai, Yuxi and He, Chao},
  journal={Light: Science \& Applications},
  volume={15},
  number={1},
  pages={254},
  year={2026},
  publisher={Nature Publishing Group UK London}
}

@article{chen2026programmable,
  title={Programmable skyrmions for communication and sensing},
  author={Chen, Long and Li, Xin Yu and Shen, Yijie and Gu, Ze and Su, Jian Lin and Xiao, Qiang and Huang, Si Qi and Qin, Shi Long and Ma, Qian and You, Jian Wei and others},
  journal={Nature Electronics},
  pages={1--10},
  year={2026},
  publisher={Nature Publishing Group UK London}
}

@article{guo2026topological,
  title={Topological robustness of classical and quantum optical skyrmions in atmospheric turbulence},
  author={Guo, Zhenyu and Peters, Cade and Mata-Cervera, Nilo and Vetlugin, Anton N and Guo, Ruixiang and Zhang, Pei and Forbes, Andrew and Shen, Yijie},
  journal={Nature Communications},
  volume={17},
  number={1},
  pages={2085},
  year={2026},
  publisher={Nature Publishing Group UK London}
}

@article{wang2024topological,
  title={Topological protection of optical skyrmions through complex media},
  author={Wang, An Aloysius and Zhao, Zimo and Ma, Yifei and Cai, Yuxi and Zhang, Runchen and Shang, Xiaoyi and Zhang, Yunqi and Qin, Ji and Pong, Zhi-Kai and Marozs{\'a}k, T{\'a}d{\'e} and others},
  journal={Light: Science \& Applications},
  volume={13},
  number={1},
  pages={314},
  year={2024},
  publisher={Nature Publishing Group UK London}
}

@article{peters2026extracting,
  title={Extracting topological features of skyrmionic light: a tutorial},
  author={Peters, Cade and Mata-Cervera, Nilo and Ornelas, Pedro and Warmbier, Robert and Shen, Yijie and Forbes, Andrew},
  journal={Advanced Photonics},
  volume={8},
  number={2},
  pages={023001--023001},
  year={2026},
  publisher={Society of Photo-Optical Instrumentation Engineers}
}

@article{zhang2024periodic,
  title={Periodic dynamics of optical skyrmion lattices driven by symmetry},
  author={Zhang, Qiang and Yang, Aiping and Xie, Zhenwei and Shi, Peng and Du, Luping and Yuan, Xiaocong},
  journal={Applied Physics Reviews},
  volume={11},
  number={1},
  year={2024},
  publisher={AIP Publishing}
}

@article{sugic2021particle,
  title={Particle-like topologies in light},
  author={Sugic, Danica and Droop, Ramon and Otte, Eileen and Ehrmanntraut, Daniel and Nori, Franco and Ruostekoski, Janne and Denz, Cornelia and Dennis, Mark R},
  journal={Nature communications},
  volume={12},
  number={1},
  pages={6785},
  year={2021},
  publisher={Nature Publishing Group UK London}
}

@incollection{Post2008,
  author = {Post, Daniel and Han, B.},
  title = {Moir\'e Interferometry},
  booktitle = {Springer Handbook of Experimental Solid Mechanics},
  publisher = {Springer},
  address = {Boston, MA},
  year = {2008},
  editor = {Sharpe, William N.},
  pages = {627--653},
  doi = {10.1007/978-0-387-30877-7_22},
  series = {Springer Handbooks}
}

@article{Beranek1982,
  author = {Beranek, Willaim and Bruinsma, Anastasius J.},
  title = {Determination of displacement and strain fields using dual-beam holographic-moir{\'e} interferometry},
  journal = {Experimental Mechanics},
  volume = {22},
  number = {9},
  pages = {317--323},
  year = {1982},
  publisher = {Springer},
  doi = {10.1007/BF02325282}
}

@article{PhysRevB.95.075420,
  title = {Twistronics: Manipulating the electronic properties of two-dimensional layered structures through their twist angle},
  author = {Carr, Stephen and Massatt, Daniel and Fang, Shiang and Cazeaux, Paul and Luskin, Mitchell and Kaxiras, Efthimios},
  journal = {Phys. Rev. B},
  volume = {95},
  issue = {7},
  pages = {075420},
  numpages = {6},
  year = {2017},
  month = {Feb},
  publisher = {American Physical Society},
  doi = {10.1103/PhysRevB.95.075420},
}

@article{Cao2018,
  author = {Cao, Yuan and Fatemi, V. and Demir, A. and Fang, S. and Tomarken, S. L. and Luo, J. Y. and Sanchez-Yamagishi, J. D. and Watanabe, K. and Taniguchi, T. and Kaxiras, E. and Ashoori, R. C. and Jarillo-Herrero, P.},
  title = {Correlated insulator behaviour at half-filling in magic-angle graphene superlattices},
  journal = {Nature},
  volume = {556},
  number = {7699},
  pages = {80--84},
  year = {2018},
  publisher = {Springer Nature},
  doi = {10.1038/nature26154}
}

@article{Cao2018b,
  author = {Cao, Yuan and Fatemi, V. and Fang, S. and Watanabe, K. and Taniguchi, T. and Kaxiras, E. and Jarillo-Herrero, P.},
  title = {Unconventional superconductivity in magic-angle graphene superlattices},
  journal = {Nature},
  volume = {556},
  number = {7699},
  pages = {43--50},
  year = {2018},
  publisher = {Springer Nature},
  doi = {10.1038/nature26160}
}

@article{Campbell2024,
  author = {Campbell, Aidan J. and Vitale, Valerio and Brotons-Gisbert, Mauro and Baek, Hyeonjun and Borel, Antoine and Ivanova, Tatyana V. and Taniguchi, Takashi and Watanabe, Kenji and Lischner, Johannes and Gerardot, Brian D.},
  title = {The interplay of field-tunable strongly correlated states in a multi-orbital moir\'e system},
  journal = {Nat. Phys.},
  volume = {20},
  number = {4},
  pages = {589--596},
  year = {2024},
  publisher = {Springer Nature},
  doi = {10.1038/s41567-024-02385-4}
}

@article{XIAO20201142,
title = {Moiré is More: Access to New Properties of Two-Dimensional Layered Materials},
journal = {Matter},
volume = {3},
number = {4},
pages = {1142-1161},
year = {2020},
issn = {2590-2385},
doi = {https://doi.org/10.1016/j.matt.2020.07.001},
author = {Yao Xiao and Jinglu Liu and Lei Fu},

}

@article{Luan2023,
  author = {Luan, H. Y. and Ouyang, Y. H. and Zhao, Z. W. and Mao, W. Z. and Ma, R. M.},
  title = {Reconfigurable moir\'e nanolaser arrays with phase synchronization},
  journal = {Nature},
  volume = {624},
  number = {7991},
  pages = {282--288},
  year = {2023},
  publisher = {Springer Nature},
  doi = {10.1038/s41586-023-06789-9}
}

@article{Hu2020,
  author = {Hu, G. and Ou, Q. and Si, G. and Wu, Y. and Wu, J. and Dai, Z. and Krasnok, A. and Mazor, Y. and Zhang, Q. and Bao, Q. and Qiu, C.-W. and Al{\`u}, A.},
  title = {Topological polaritons and photonic magic angles in twisted $\alpha$-MoO$_3$ bilayers},
  journal = {Nature},
  volume = {582},
  number = {7811},
  pages = {209--213},
  year = {2020},
  publisher = {Springer Nature},
  doi = {10.1038/s41586-020-2359-9}
}

@article{10.1021/acs.nanolett.0c01673,
    author = {Duan, Jiahua and Capote-Robayna, Nathaniel and Taboada-Gutiérrez, Javier and Álvarez-Pérez, Gonzalo and Prieto, Iván and Martín-Sánchez, Javier and Nikitin, Alexey Y. and Alonso-González, Pablo},
    title = {Twisted Nano-Optics: Manipulating Light at the Nanoscale
with Twisted Phonon Polaritonic Slabs},
    journal = {Nano Letters},
    volume = {20},
    number = {7},
    pages = {5323-5329},
    year = {2020},
    month = {06},

    issn = {1530-6984},
    doi = {10.1021/acs.nanolett.0c01673},
}

@article{Wang2026,
  author = {Wang, M. and Lv, N. and Zhang, Z. and Wang, P. and Chen, Y. and Li, X. and Wang, Y. and Maier, S. A. and Zhang, X. and Li, Z.},
  title = {Chiral orbital lasing in a twisted bilayer metasurface},
  journal = {Nat. Commun.},
  volume = {17},
  pages = {2369},
  year = {2026},
  publisher = {Springer Nature},
  doi = {10.1038/s41467-026-69665-w}
}

@article{PhysRevApplied,
  title = {Twistronics-based polarization engineering},
  author = {Orazbay, Mendikhan and Valagiannopoulos, Constantinos},
  journal = {Phys. Rev. Appl.},
  volume = {22},
  issue = {5},
  pages = {054028},
  numpages = {13},
  year = {2024},
  month = {Nov},
  publisher = {American Physical Society},
  doi = {10.1103/PhysRevApplied.22.054028},
}

@article{WU2025113833,
title = {Photon Moiré lattices: light field localization and bandgap engineering for theoretical and technological breakthroughs in low-threshold, high-power lasers},
journal = {Optics \& Laser Technology},
volume = {192},
pages = {113833},
year = {2025},
issn = {0030-3992},
doi = {https://doi.org/10.1016/j.optlastec.2025.113833},
author = {Zibo Wu and Yue Song and Jishun Liu and Hongbo Sha and Yongyi Chen and Hao Zhang and Mengjie Shi and Li Qin and Lei Liang and Peng Jia and Cheng Qiu and Yuxin Lei and Yubing Wang and Lijun Wang},
}

@article{Shi2026,
  author = {Shi, P. and Gou, X. and Zhang, Q. and Wang, Y. and Liu, S. and Chen, L. and Li, Z. and Zhang, H. and Fang, Y. and Sun, Z.},
  title = {Observation of strong spin-orbit couplings in plasmonic spin-twistronics topological lattices},
  journal = {Nat. Commun.},
  volume = {17},
  pages = {1905},
  year = {2026},
  publisher = {Springer Nature},
  doi = {10.1038/s41467-026-68629-4}
}

@article{Du2023Science,
  author = {Du, Luojun and Molas, Maciej R. and Huang, Zhiheng and Zhang, Guangyu and Wang, Feng and Sun, Zhipei},
  title = {Moir\'e photonics and optoelectronics},
  journal = {Science},
  volume = {379},
  number = {6639},
  pages = {eadg0014},
  year = {2023},
  doi = {10.1126/science.adg0014}
}

@article{Oudich2024,
  author = {Oudich, Mourad and Kong, Xianghong and Zhang, Tan and Qiu, Chengwei and Jing, Yun},
  title = {Engineered moir\'e photonic and phononic superlattices},
  journal = {Nature Materials},
  volume = {23},
  pages = {1169--1178},
  year = {2024},
  doi = {10.1038/s41563-024-01950-9}
}

@article{skyrmion2018,
author = {S. Tsesses  and E. Ostrovsky  and K. Cohen  and B. Gjonaj  and N. H. Lindner  and G. Bartal },
title = {Optical skyrmion lattice in evanescent electromagnetic fields},
journal = {Science},
volume = {361},
number = {6406},
pages = {993-996},
year = {2018},
doi = {10.1126/science.aau0227},

}

@article{Shen2024,
  author = {Shen, Yijie and Zhang, Qiang and Shi, Peng and Du, Luping and Yuan, Xiaocong and Zayats, Anatoly V.},
  title = {Optical skyrmions and other topological quasiparticles of light},
  journal = {Nat. Photon.},
  volume = {18},
  number = {1},
  pages = {15--25},
  year = {2024},
  publisher = {Springer Nature},
  doi = {10.1038/s41566-023-01325-7}
}

@article{Schwab2025,
  author = {Schwab, J. and Neuhaus, A. and Dreher, P. and Mayer, B. and Seidel, F. and Evers, J. and Röhlsberger, R. and Bock, V. and von Freymann, G. and Wrobel, F. and Decker, M.},
  title = {Skyrmion bags of light in plasmonic moir\'e superlattices},
  journal = {Nat. Phys.},
  volume = {21},
  pages = {988--994},
  year = {2025},
  publisher = {Springer Nature},
  doi = {10.1038/s41567-025-02873-1}
}

@article{tunable2026,
author = {Schwab, Julian and Mangold, Florian and Frank, Bettina and Davis, Timothy J. and Giessen, Harald},
title = {Tunable Skyrmion, Meron, and Skyrmion Bag Textures in Surface Phonon Polariton Lattices},
journal = {Nanophotonics},
volume = {15},
number = {4},
pages = {e70015},
doi = {https://doi.org/10.1002/nap2.70015},

year = {2026}
}

@article{trilayer2025,
author = {Schwab, Julian and Mangold, Florian and Frank, Bettina and Davis, Timothy J. and Giessen, Harald},
title = {Skyrmion bag robustness in plasmonic bilayer and trilayer moiré superlattices},
journal = {Nanophotonics},
volume = {14},
number = {23},
pages = {3955-3964},
doi = {https://doi.org/10.1515/nanoph-2024-0581},

year = {2025}
}

@article{dev,
author  = {Dev, Vasu and Shen, Yijie},
  title   = {Freespace ``twistronics'': moir\'e structured light for optical supertopologies},
  journal = {Light: Science \& Applications},
  volume  = {15},
  pages   = {396},
  year    = {2026},
  doi     = {10.1038/s41377-026-02533-w}
}

@article{shen2023,
author = {Yijie Shen and Bingshi Yu and Haijun Wu and Chunyu Li and Zhihan Zhu and Anatoly V. Zayats},
journal = {Advanced Photonics},
number = {11},
pages = {015001},
publisher = {},
title = {Topological transformation and free-space transport of photonic hopfions},
volume = {5},
month = {1},
year = {2023},

}

@article{hh5s-cprt,
  title = {Space-Time Optical Hopfion Crystals},
  author = {Lin, Wenbo and Mata-Cervera, Nilo and Ota, Yasutomo and Shen, Yijie and Iwamoto, Satoshi},
  journal = {Phys. Rev. Lett.},
  volume = {135},
  issue = {8},
  pages = {083801},
  numpages = {6},
  year = {2025},
  month = {Aug},
  publisher = {American Physical Society},
  doi = {10.1103/hh5s-cprt},

}

@article{Whitehead1947,
  author = {Whitehead, J. H. C.},
  title = {An Expression of {Hopf}'s Invariant as an Integral},
  journal = {Proceedings of the National Academy of Sciences of the United States of America},
  volume = {33},
  number = {5},
  pages = {117--123},
  year = {1947},
  doi = {10.1073/pnas.33.5.117},
  pmid = {16578254},
  pmcid = {PMC1079004}
}

@article{Linking,
  title = {Linking Numbers, Spin, and Statistics of Solitons},
  author = {Wilczek, Frank and Zee, A.},
  journal = {Phys. Rev. Lett.},
  volume = {51},
  issue = {25},
  pages = {2250--2252},
  numpages = {0},
  year = {1983},
  month = {Dec},
  publisher = {American Physical Society},
  doi = {10.1103/PhysRevLett.51.2250},
 
}

@misc{sound,
      title={Twistsonics: engineering acoustic topological textures in moire sound lattices}, 
      author={Wen-Yu Wang and Shuai Liu and Xiang-Yuan Xu and Qing Tong and Hao Ge and Yijie Shen and Ming-Hui Lu},
      year={2026},
      eprint={2608.01278},
      archivePrefix={arXiv},
      primaryClass={physics.app-ph},
     
}

@article{Ochoa2021DegradationOP,
  title={Degradation of Phonons in Disordered Moir{\'e} Superlattices.},
  author={H{\'e}ctor Ochoa and Rafael M. Fernandes},
  journal={Physical review letters},
  year={2021},
  volume={128 6},
  pages={
          065901
        },

}

@article{Wang2025,
  author = {Wang, B. and Che, Z. and Cheng, C. and Chen, Y. and Liu, Y. and Zhang, X. and Wang, Z. and Shen, Y.},
  title = {Topological water-wave structures manipulating particles},
  journal = {Nature},
  volume = {638},
  pages = {394--400},
  year = {2025},
  publisher = {Springer Nature},
  doi = {10.1038/s41586-024-08384-y}
}

@article{CrX3,
    author = {Akram, Muhammad and LaBollita, Harrison and Dey, Dibyendu and Kapeghian, Jesse and Erten, Onur and Botana, Antia S.},
    title = {Moir\`e  Skyrmions and Chiral Magnetic Phases in Twisted CrX3 (X = I, Br, and Cl) Bilayers},
    journal = {Nano Letters},
    volume = {21},
    number = {15},
    pages = {6633-6639},
    year = {2021},
    month = {08},
    issn = {1530-6984},
    doi = {10.1021/acs.nanolett.1c02096},
}

\end{document}